\documentclass{webofc}
\usepackage[varg]{txfonts}   % Web of Conferences font
\usepackage{comment}
\excludecomment{confidential}
\begin{document}
\title{Insights on strange quark hadronization in small collision system with ALICE: multiple strange hadrons and $\Sigma^{\pm}$ baryons}

\author{\firstname{\textit{Sara}} \lastname{Pucillo}\inst{1}\fnsep\thanks{\email{sara.pucillo@cern.ch}}  for the ALICE Collaboration }

\institute{Dipartimento di Fisica, Università degli Studi di Torino, and Sezione INFN Torino}

\abstract{%Among the most iconic results of Run 1 and Run 2 of the LHC is the observation of enhanced production of (multi-)strange to non-strange particles, gradually rising from low-multiplicity to high-multiplicity pp or p-Pb collisions and reaching values close to those measured in peripheral Pb--Pb collisions. More insightful information about the production mechanism could be provided by measuring the full Probability Density Function (PDF) for the production of each strange particle specie and investigating if any deviation from pure uncorrelated statistical behavior is observed. Using this novel method, we can determine whether strangeness enhancement is connected to the high-multiplicity tail of the PDF or to a progressive increase in the number of events with few strange particles produced. In this contribution, we present new results on the full PDF for the production of $K^{0}_{S}$, $\Lambda$, $\Xi^{-}$ and $\Omega^{-}$ in pp collisions at $\sqrt{s}$ = 5.02 TeV as a function of the multiplicity. In addition, we present new results on the transverse momentum spectra of $\Sigma^{+}$ and its charge conjugate anti-particle, in both minimum bias and high-multiplicity triggered pp collisions at $\sqrt{s}$ = 13 TeV. The results are compared to state-of-the-art phenomenological models implemented in commonly-used Monte Carlo event generators, drastically enhancing the sensitivity to the different processes implemented in each approach. \textbf{questo era quello sul sito della conferenza}
Among the most iconic results of Run-1 and Run-2 of the LHC is the observation of enhanced production of (multi-)strange to non-strange particles, gradually rising from low-multiplicity to high-multiplicity pp or p--Pb collisions and reaching values close to those measured in peripheral Pb--Pb collisions \cite{nature}. The observed behaviour cannot be quantitatively reproduced by any of the available QCD-inspired MC generators. In this contribution two extensions of this study are presented: the measurement of $\Sigma$ baryons and the first measurement of the full Probability Density Function (PDF) for $K^{0}_{S}$, $\Lambda$, $\Xi^{-}$ and $\Omega^{-}$, therefore extending the study of strangeness production beyond the average of the distribution. This novel method represents a unique opportunity to test the connection between charged and strange particle multiplicity production.}
\maketitle
\section{Introduction} \label{intro}
Among several probes of the Quark-Gluon Plasma (QGP) formation, the so-called \textit{Strangeness Enhancement} (SE) was one of the first proposed \cite{rafelski} and observed experimentally \cite{na57}. In addition, to confirm the SE at the highest center-of-mass energy \cite{PbPb2.76,pPbV0,pPbCasc}, the ALICE Collaboration carried out a comprehensive study of strange hadron production (relative to pions) as a function of the charged-particle multiplicity \cite{nature}. The main finding is that strangeness enhances progressively with multiplicity, across different colliding systems and center-of-mass energies \cite{nature,pp7,pp13}. The invariance of the SE pattern on the colliding system suggested that the mechanisms at play in high-multiplicity pp interactions could be the same as those involved in particle formation in Pb--Pb collisions. Moreover the increase, previously unexpected in pp interactions, is proportional to the strangeness content being the highest for the triple-strange $\Omega$. However, the observed behaviour cannot be quantitatively reproduced by any of the available QCD-inspired models, suggesting that further developments are needed to obtain a complete microscopic understanding of strangeness production.

\section{$\Sigma$ baryon production} \label{sigma}
\begin{figure}
    \centering
    \includegraphics[width=0.5 \textwidth, height=0.35\linewidth]{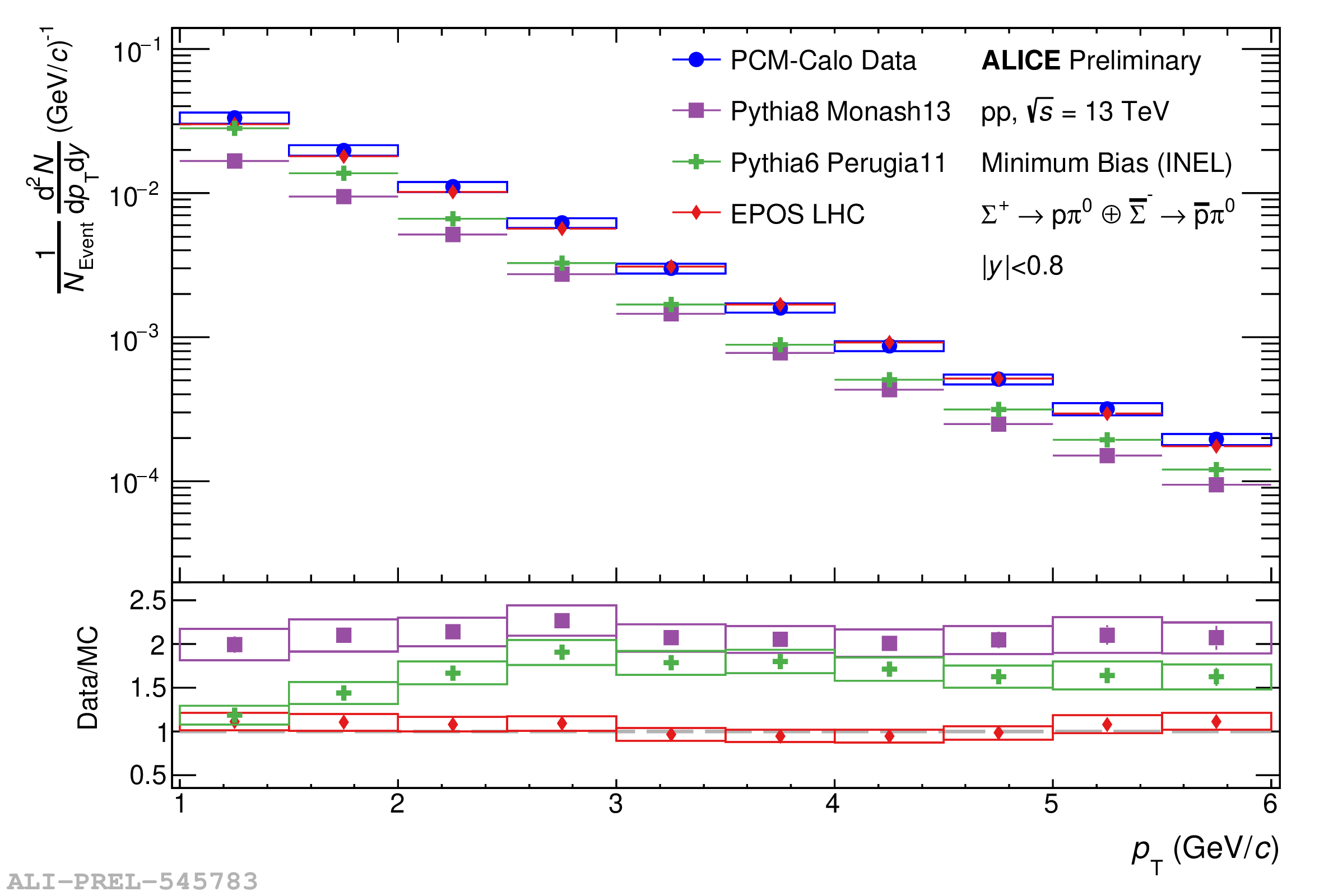} 
    \hspace*{0.35cm}
    \includegraphics[width=0.4 \textwidth, height=0.4\linewidth]{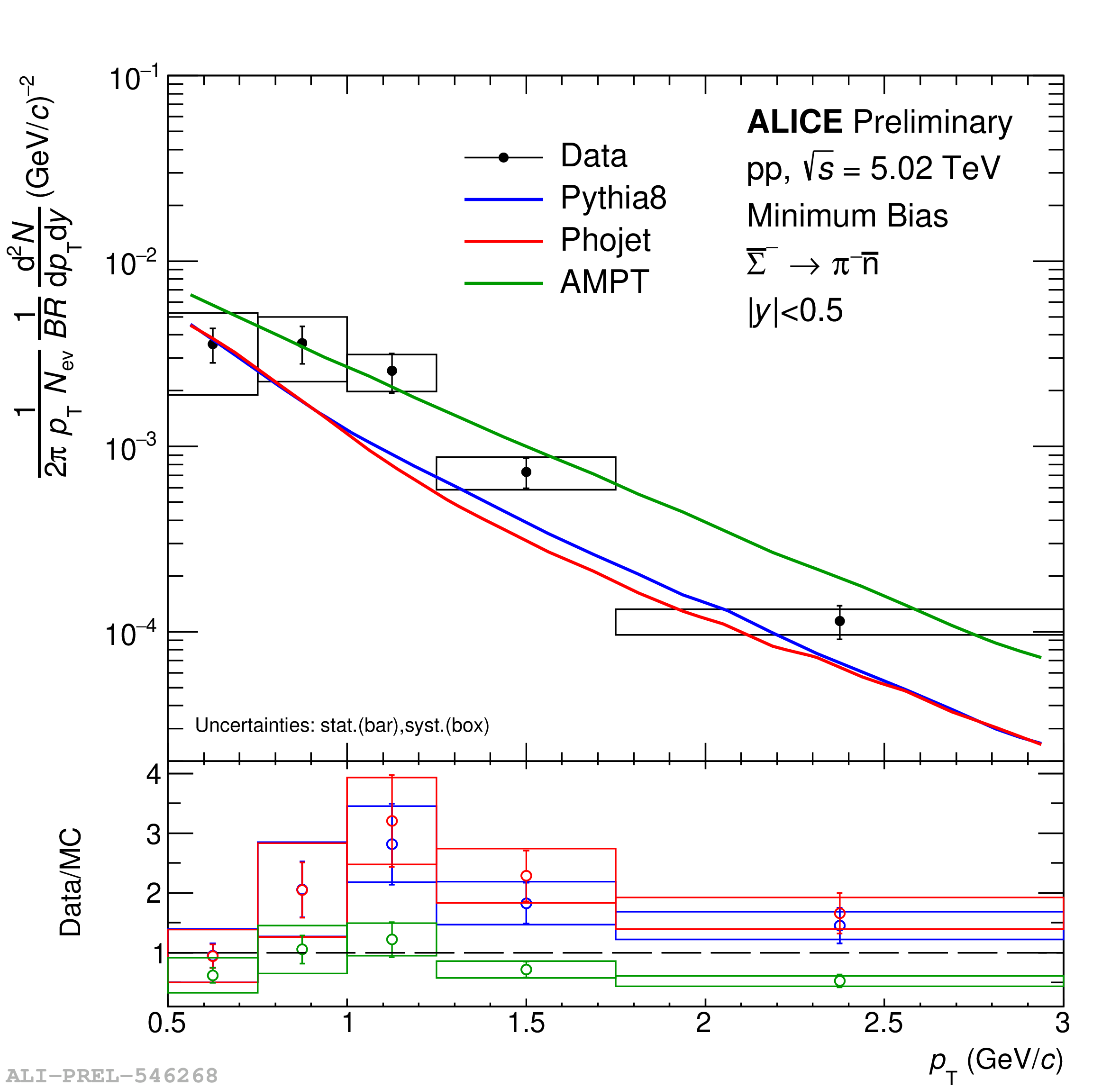}
    \caption{\textit{Left}: $p_{T}$ spectrum obtained for $\Sigma^{+}$ and its charge conjugate anti-particle in pp collisions at $\sqrt{s}$ = 13 TeV. \textit{Right}: $p_{T}$ spectrum obtained for $\overline{\Sigma}^{-}$ in pp collisions at $\sqrt{s}$ = 5.02 TeV.}
    \label{fig:sigma}
\end{figure}

The measurement of the $\Sigma$ hyperon production is an important addition to what shown in \cite{nature} since it has the same strangeness content of $\Lambda$ but it is charged. The ALICE Collaboration investigated the $\Sigma^{+} \rightarrow p + \pi^{0}$ decay mode (branching ratio of 51.57\%) in pp collisions at $\sqrt{s}$ = 13 TeV by reconstructing the $\pi^{0} \rightarrow \gamma + \gamma$ using two independent reconstruction methods \cite{photon}: in the first one, each photon is reconstructed from conversion $\gamma \rightarrow e^{+}e^{-}$ (PCM), while in the second one, one photon is reconstructed from conversion and one is measured in the calorimeters (PCM-Calo). On the left of Fig. \ref{fig:sigma} the $p_{T}$ spectrum obtained for $\Sigma^{+}$ and its charge conjugate anti-particle is shown and compared with different models. EPOS LHC \cite{EPOS} and Pythia8 \cite{Pythia} are both able to reproduce the $p_{T}$ shape, but only EPOS LHC describes the yields satisfactorily. \newline
The Collaboration also measured for the first time the $\overline{\Sigma}$ production in pp collisions at $\sqrt{s}$ = 5.02 TeV reconstructing the decay into a anti-neutron (by means of the PHOton Spectrometer calorimeter \cite{phos}) and a charged pion ($\overline{\Sigma}^{-} \rightarrow \pi^{-} + \overline{n}$, branching ratio of 99.848\%). On the right of Fig. \ref{fig:sigma} the $p_{T}$ spectrum for $\overline{\Sigma}^{-}$ obtained with this very promising new technique is shown and compared to phenomenological models. 

\section{Multiple strange hadron production} \label{PDF}
\begin{figure}
    \centering
    \includegraphics[width=0.45 \textwidth, height=0.45\linewidth]{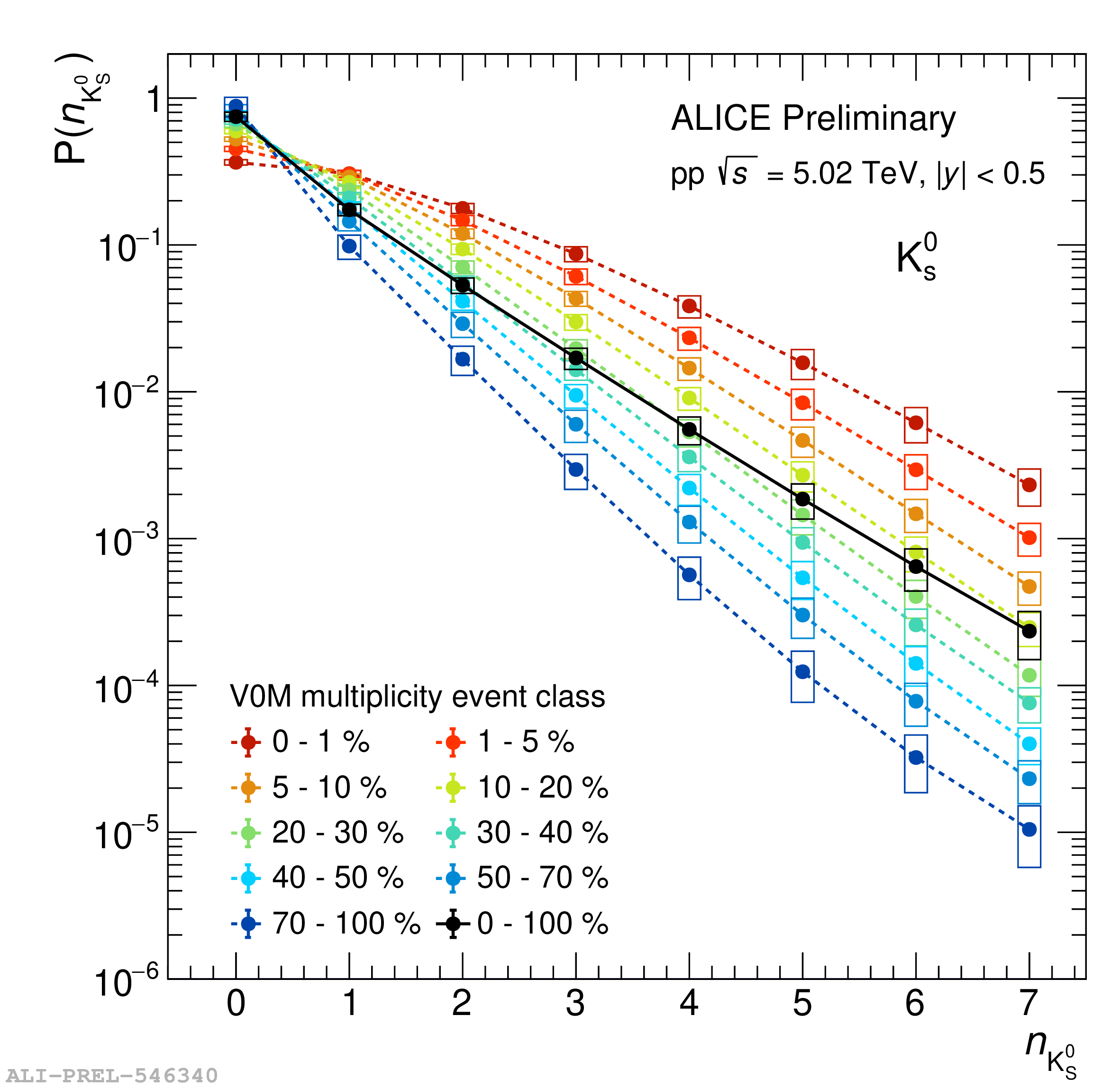} 
    \hspace*{0.3cm}
    \includegraphics[width=0.45 \textwidth, height=0.45\linewidth]{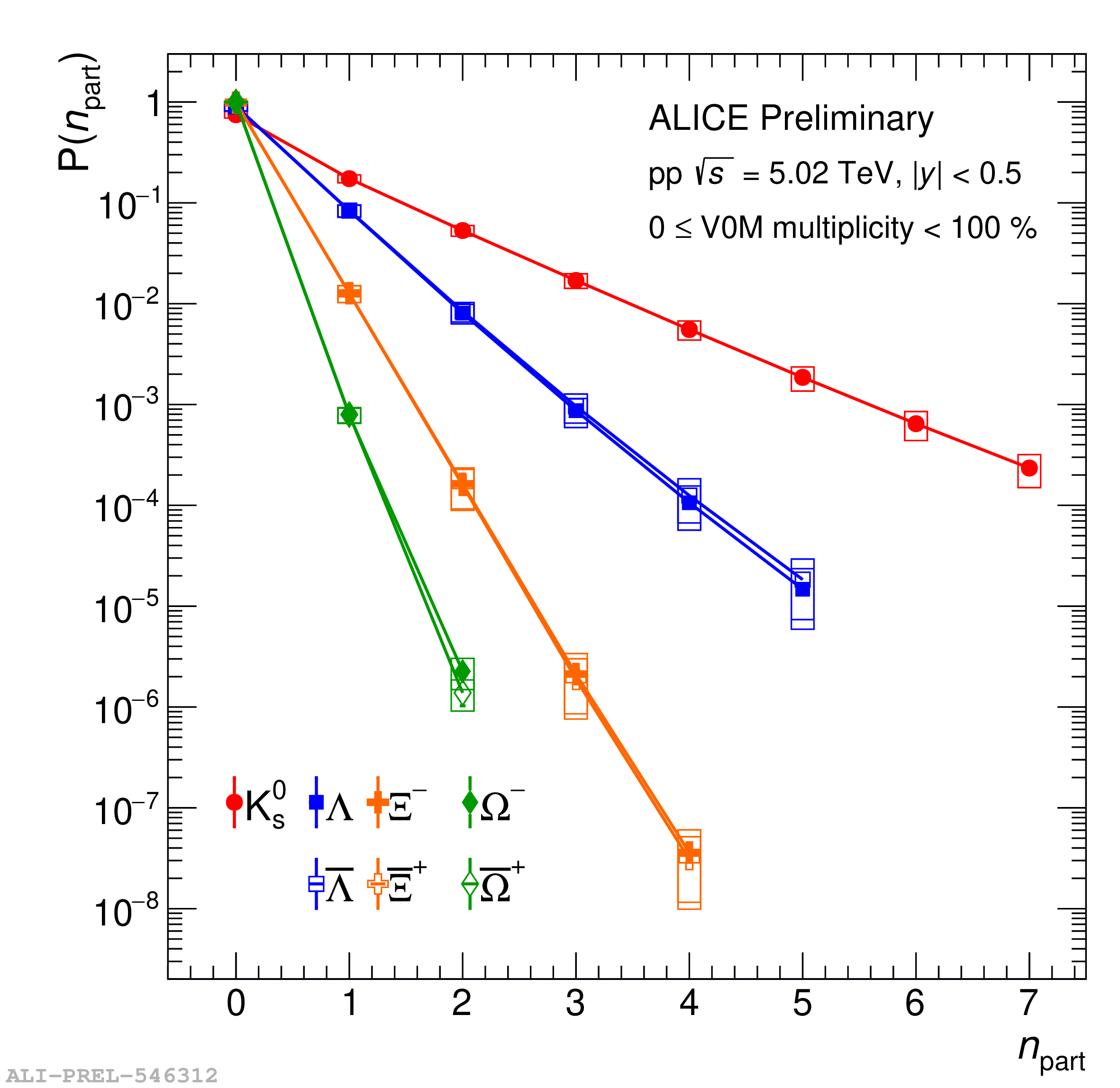}
    \caption{$Left$: The probability to produce $n_{K^{0}_{S}}$ per event for several multiplicity classes. $Right$: P($n_{K^{0}_{S}}$), P($n_{\Lambda}$) (P($n_{\overline{\Lambda}}$)), P($n_{\Xi^{-}}$) (P($n_{\overline{\Xi}^{+}}$)) and P($n_{\Omega^{-}}$) (P($n_{\overline{\Omega}^{+}}$)) for the integrated multiplicity (0-100 \%). Results obtained for particles have been reported as full points, while for the antiparticles as open ones. In both figures the lines are not fit, they connect only data points.}
    \label{fig:k0s_all}
\end{figure}

In order to extend all the above observations to the measurement of the full PDF a new technique based on counting the number of strange particles event-by-event in pp collisions at $\sqrt{s}$ = 5.02 TeV has been adopted.%, exploiting the high-statistics data set collected by ALICE at the LHC during the 2017 pp data taking campaign.
\newline
Signal extraction has been performed investigating a procedure based on signal/background weights obtained from the data after the application of selective cuts on topological features of the weak decay and particle identification. As a first step, a fit to the 1-dimensional invariant mass spectrum given by the sum of a function for the signal and one for the background was performed in different $p_{T}$ and multiplicity bins. In this way, for each invariant mass it is possible to define a probability that the candidate is signal (ratio between the value of the signal function and the total one) or background (ratio between the value of the background function and the total one). After this first step, all events are re-analysed in order to apply the procedure to all N candidates per event. Event by event, each candidate is considered, associating the probability for it to be signal or background (from previous step) which depends on the invariant mass and transverse momentum. Then, all the products of the different probabilities associated to each candidate were computed and were grouped according to the total signal yield. Consequently, it is possible to obtain that all candidates are signal, are background, or all the intermediate situations, having for each event with N candidates a full probability spectrum spanning from 0 to N. Summing the result obtained for each event, it was possible to measure the reconstructed spectra for each multiplicity class. 
\begin{confidential}
Using this method it is possible to define a probability that the candidate is signal or background fitting the 1D invariant mass spectrum in different $\textit{p}_{T}$ and multiplicity bins. Applying the procedure to all N candidates per event and combing the probability associated to each candidate it is possible to obtain the probability that all candidates are signal, are background, or all the intermediate situations. For each event with N candidates a full probability spectrum spanning from 0 to N is obtained and summing what was obtained for each event it was possible to measure the reconstructed spectra for each multiplicity class. 
\end{confidential}
To estimate the correction for the detector response, MC productions with realistic strange particle $\textit{p}_{T}$ distributions and detector conditions have been used to perform a uni-dimensional bayesian unfolding procedure \cite{unf}, leading to the corrected PDF for $K^{0}_{S}$, $\Lambda$, $\Xi^{-}$ and $\Omega^{-}$.\newline
Results are reported in Fig. \ref{fig:k0s_all}, for $K^{0}_{S}$ in several multiplicity classes (\textit{left}) and for all the analyzed particles for the integrated multiplicity (\textit{right}). The probability to produce more than one particle per event (e.g. 2) increases with the event charged-particle multiplicity moving from blue to red points consistently across the full PDF. \newline
This result represents a unique opportunity to test the connection between charged and strange particle multiplicity production all the way to very "extreme" situations, e.g. spanning from events with 7 $K^{0}_{S}$ at low average multiplicity -- where $<dN/d\eta>_{|\eta|<0.5} \sim 3$, with potential large fluctuations -- to events with 0 $K^{0}_{S}$ at high multiplicity -- where $<dN/d\eta>_{|\eta|<0.5} \sim 20$. %Note that, in each V0M bin, the multiplicity can fluctuate and the average can significantly change for events with small or large number of $K^{0}_{S}$ for example.

\subsection{Average probability for the production yields} \label{yields}
From the measurement of P($n_{part}$) it is possible to calculate the average probability \newline $<Y_{n-part}>$ for the production of $n$ particles using the Eq. \ref{eq:yields} that contains the combinatorial factor and the probability to produce $i-$particles per event of a given species (the i-th bin of the PDF for the particle of interest). %Note that in the formula the sum does not really extend to infinity being that it was possible to measure the probability to produce $n$ particles up to a specific number of particles (7 for $K^{0}_{S}$, 5 for $\Lambda (\overline{\Lambda})$, 4 for $\Xi^{-} (\overline{\Xi}^{+})$ and 2 for $\Omega^{-} (\overline{\Omega}^{+})$).

\begin{equation}
  <Y_{n-part}> = \sum_{i=0}^{\infty} \frac{i!}{n!(i-n)!}Y_{i-part}.
  \label{eq:yields}
\end{equation}

\begin{figure}
    \centering
    \includegraphics[width=0.45 \textwidth, height=0.45\linewidth]{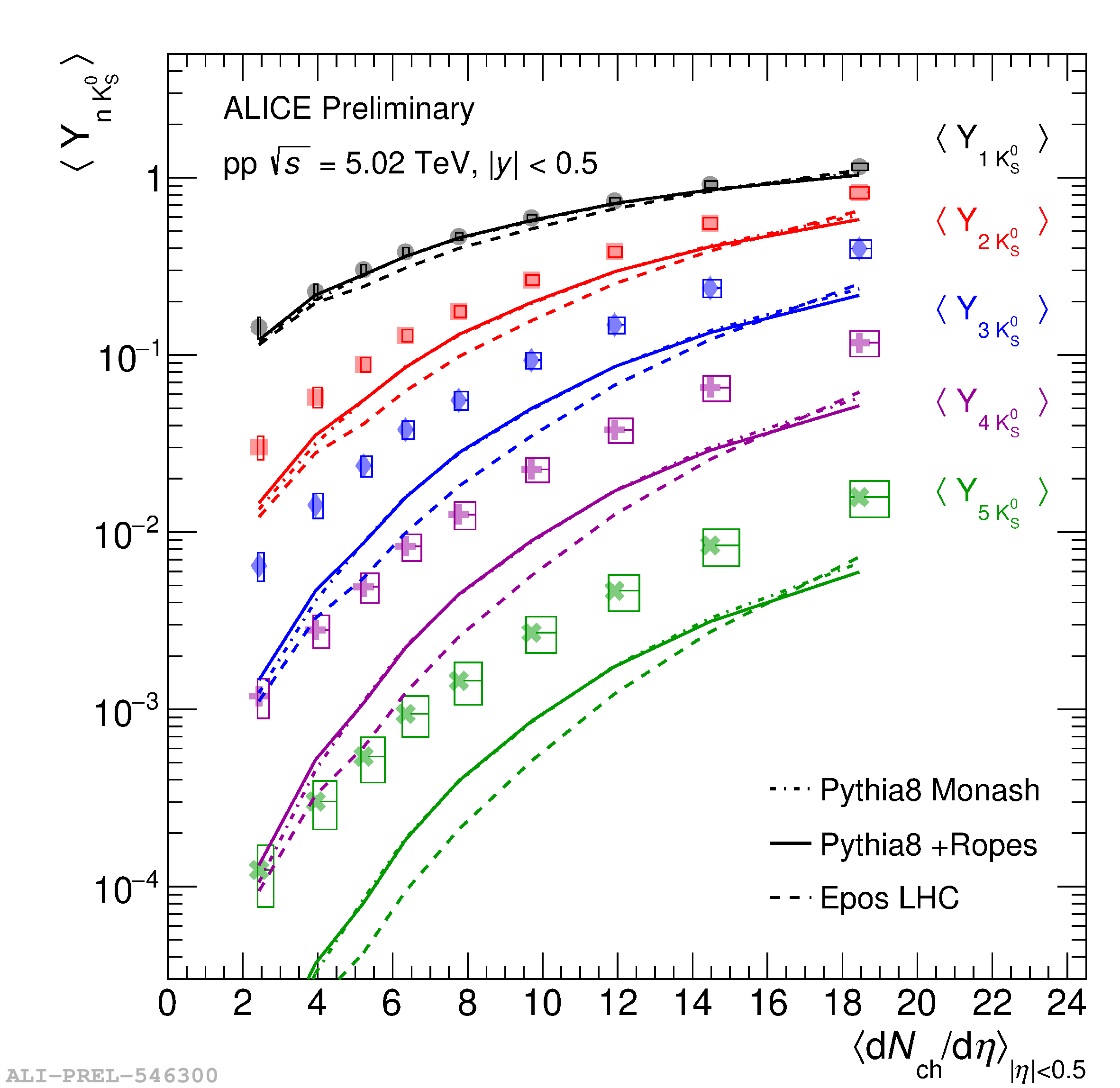} 
    \hspace*{0.3cm}
    \includegraphics[width=0.45 \textwidth, height=0.45\linewidth]{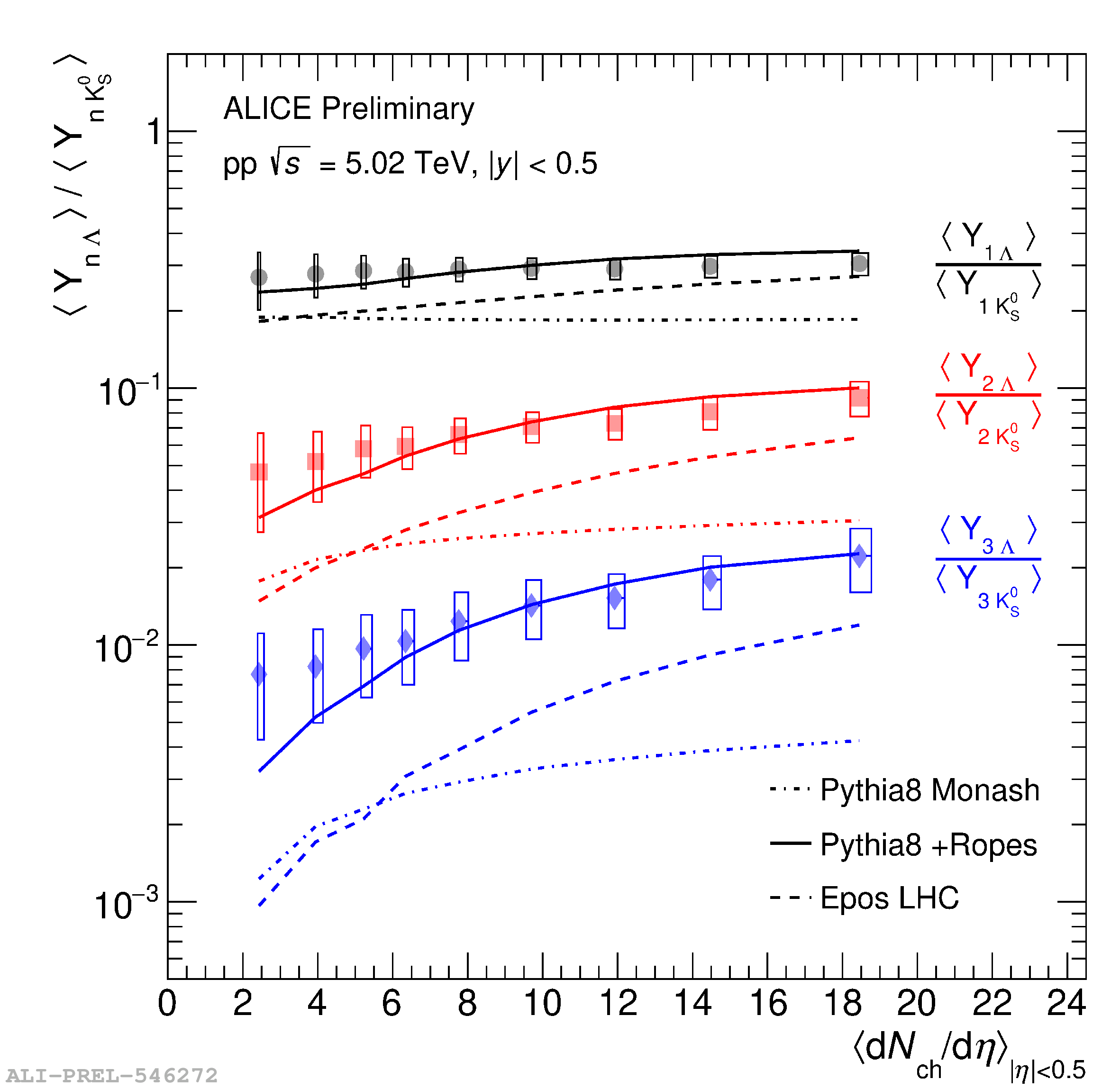}
    \caption{$\textit{Left}$: Average production yield of 1, 2, 3, 4 and 5 $K^{0}_{S}$ respectively in black, red, blue, orange and green, as a function of the charged particle multiplicity compared with Pythia8 Monash \cite{Pythia}, Pythia8 + Ropes (with QCD-CR) and EPOS LHC \cite{EPOS} models. $\textit{Right}$: Ratio $<Y_{n\Lambda}> / <Y_{nK^{0}_{S}}>$ as a function of the charged particle multiplicity compared with Pythia8 Monash \cite{Pythia}, Pythia8 + Ropes (with QCD-CR) \cite{PythiaRopes} and EPOS LHC \cite{EPOS} models.}
    \label{fig:yields}
\end{figure}

%$<Y_{n-part}>$ represents the average yields of producing 1 part/ev, particle doublets, triplets, etc.. . In order to calculate $<Y_{2-part}>$  ($<Y_{2-part}>$  = $Y_{2-part}$  + 3 $Y_{3-part}$ + 6 $Y_{4-part}$ +..) it is necessary to take into account all the possible combinations to have a doublet. So if there are 2 candidates ($Y_{2-part}$), there is only one chance of getting a doublet of particles (combinatorial factor 1), if there are 3 candidates (e.g called ABC, $Y_{3-part}$) there are different possible combinations that allow to have 2 particles/ev (e.g AB,AC,BC), hence combinatorial factor 3 and so on. \newline
\noindent In Fig. \ref{fig:yields}, the average production yield of 1, 2, 3, 4 and 5 $K^{0}_{S}$ as a function of the charged particle multiplicity are reported on the left. The increase with multiplicity is more than linear for the production of multiple strange hadrons. Comparing the results with Pythia8 Monash \cite{Pythia}, Pythia8 + Ropes (with QCD-Color Reconnection) \cite{PythiaRopes} and EPOS LHC \cite{EPOS}, one can see that the agreement worsens as the number of particles per event increases.\newline
Moreover, the production yields of more than one particle allow to evaluate the ratio \newline $<Y_{n\Lambda}> / <Y_{nK^{0}_{S}}>$ as a function of the multiplicity which is a very important quantity to disentangle baryon from strangenes-related effects in the multiplicity-dependent enhancement pattern. This is reported in the right side of Fig. \ref{fig:yields}, where an increasing pattern is observed when looking at multiple strange particle production. This increasing pattern is rather well reproduced by Pythia8 + Ropes (with QCD-CR), where specific mechanisms are implemented, which lead to an effective baryon enhancement. All these observations suggest that in every strange-hadron/$\pi$ \cite{nature, pp13} plots as a function of multiplicity the observed enhancement can be partly attributed to strangeness and partly to baryon number.

%
% BibTeX or Biber users please use (the style is already called in the class, ensure that the "woc.bst" style is in your local directory)
% \bibliography{name or your bibliography database}

\end{document}